\documentclass[12pt,aps,prb,floats,graphicx,euscript,amsmath] {revtex4}
\usepackage[cp1251]{inputenc}
\usepackage[dvips]{graphicx}
\DeclareMathAlphabet\EuScript{U}{eus}{m}{n}
\SetMathAlphabet\EuScript{bold}{U}{eus}{b}{n}

\renewcommand{\min}{\mathop{\rm min}\nolimits}

\def\lapprox{\,\raise0.4ex\hbox{$<$}\kern-0.8em\lower0.7ex\hbox{$\sim$}\,}
\def\gapprox{\,\raise0.4ex\hbox{$>$}\kern-0.8em\lower0.7ex\hbox{$\sim$}\,}
\begin{document}
\bibliographystyle{prsty}
\title{Double-exciton component of the cyclotron spin-flip mode in a quantum Hall ferromagnet}
\author{S. Dickmann and V.M. Zhilin}
\affiliation{Institute for Solid State Physics of RAS, Chernogolovka
142432, Moscow District, Russia. }
\begin{abstract}
\vspace{0.mm}
We report on the calculation of the cyclotron spin-flip excitation (CSFE) in a spin-polarized quantum Hall system at unit filling. This mode has a double-exciton component which contributes to the CSFE correlation energy but can not be found by means of a mean field (MF) approach. The result is compared with available experimental data.

\noindent
PACS numbers 73.21.Fg, 73.43.Lp, 78.67.De
\end{abstract}
\maketitle

\bibliographystyle{prsty}
A two-dimensional electron gas (2DEG) in a high perpendicular
magnetic field possesses many remarkable features.\cite{an82} In particular, it presents a rare
case of strongly correlated system governed by {\em real} Coulomb interaction (not by a model
Hamiltonian!) where, nevertheless, some solutions of the quantum many-body problem can be found
exactly.  Indeed, under the conditions of integer  quantum Hall effect (when the filling factor
is $\nu=1,2,3,...$), the one-cyclotron magnetoplasma and the lowest spin-flip modes are
calculated analytically to the leading order in the parameter $r_{\rm c}\!=\!E_{\rm
C}/\hbar\omega_c$. \cite{by81,by83,ka84,di05} [$\omega_c$ is the cyclotron frequency; $E_{\rm
C}\!=\alpha e^2/\kappa l_B$ is the characteristic interaction energy, $\alpha$ being the average
form-factor related to the finite thickness of the 2DEG ($0.3\lapprox\alpha\!<\!1$); $l_B$ is
the magnetic length.] This astounding property is the feature of either filled or half-filled
highest-occupied Landau level (LL) where the simplest-type excitations are single excitons or
superposition of single-exciton modes. The many-body problem is thereby reduced to the two-body
one, i.e. to the interaction of electron with an effective hole. Being quite
in the context of similar studies, the present letter concerns however the {\it case which can not be reduced to a
single-exciton problem}.

We remind that 2DEG excitons are characterized by sublevels $a\!=\!(n_a,\sigma_a)$ and
$b\!=\!(n_b,\sigma_b)$ where electron is promoted from the $n_a$-th LL with spin-component
$S_z\!=\!\sigma_a$ to the $n_b$-th LL with $S_z\!=\!\sigma_b$. The relevant quantum numbers are
$\delta n\!=\!n_b\!-\!n_a$, $\delta S_z\!=\!\sigma_b\!-\!\sigma_a$, and the two-dimensional (2D)
wave vector ${\bf q}$. The single exciton problem is exactly solvable in the following cases:
(i) at odd filling $\nu$ when $\delta n\!=\!1$ and $\delta S_z\!=\!0$ (magnetoplasmon) or
$\delta n\!=\!0$ and $\delta S_z\!=\!-1$ (spin wave);\cite{by81,ka84,dizh05} (ii) at even $\nu$
when $\delta n\!=\!1$ and $\delta S_z\!=\!0,\pm 1$ (magnetoplasmon and spin-flip
triplet).\cite{ka84,di05,dizh05} At the same time the two-body problem may be discussed within
a MF approach (in some publications called `time-dependent Hartree-Fock'
approximation$\,$\cite{mc85,lo93}) which excludes any quantum fluctuations from a single exciton
to double- or many-exciton states. For the above simplest cases of $\delta n$ and $\delta S_z$,
the MF calculation gives an asymptotically exact result which may be found perturbatively to the
first order in $r_{\rm c}$,\cite{foot1} because these $(\delta n,\,\delta S_z)$ sets can not
correspond to any states except single-exciton modes. Any complication of $(\delta n,\delta
S_z)$ makes the calculations substantially more difficult due to the necessary expansion of the
basis to the entire continuous set of many-exciton states with the same total numbers $\delta
n$, $\delta S_z$, and ${\bf q}$. For example, the double-cyclotron plasmon with $\delta
n\!=\!2$, $\delta S_z\!=\!0$ and with given ${\bf q}$ `dissociates' into double-exciton states
consisting of one-cyclotron plasmon's pairs with the total momentum equal to ${\bf
q}$.\cite{ka84} At odd $\nu$, a similar `dissociation' occurs for the CSFE, where $\delta n\!=\!-\delta S_z\!=\!1$. The proper double-exciton states are pairs of a
magnetoplasmon ($\delta n\!=\!1,\,\delta S_z\!=\!0$) and a spin wave ($\delta
n\!=\!0,\,\delta S_z\!=\!-1$). The problem thus changes from the two-body case to the {\it
four-body} one, and the correct solution should be presented in the form of combination of the
single-exciton mode and continuous set of double-exciton states.\cite{dizh05} It is important
that in both cases the desired solution corresponds to a discrete line against the background of a
continuous spectrum of free exciton pairs. The technique of correct solution has to be of
essentially non-Hartree-Fock (non-HF) type. Actually this letter concerns the fundamental question of
consistency of the MF approach.

By considering the case of unit filling factor where the number of electrons is equal to the
number of magnetic flux quanta $N_\phi$, now we report on a study of the CSFE with ${\bf q}\!=\!0$. This state is optically active and identified in the ILS experiments.\cite{pi92,va06} Besides, it is exactly
this spin-flip magnetoplasma mode which is the key component of the elementary perturbation used
in the microscopic approach to the skyrmionic problem.\cite{di02} The calculation is performed
in `quasi-analytical' way which should, in principle, lead to the result which is exact in the
leading approximation in $r_{\rm c}$. In our case the envelope function determining the
combination of the double-exciton states is one-dimensional --- i.e., it only depends on the
modulus of the excitons' relative momentum. This function is chosen in the form of expansion
over infinite orthogonal basis, where every basis vector obeys a specific symmetry condition
necessary for the total envelope function. Even to the {\em first-order approximation} in
$r_{\rm c}$, we obtain a non-HF correction to the former HF result$\,$\cite{pi92,lo93} for the
CSFE energy.

As a technique, we use the {\em excitonic representation} (ER) which is a convenient tool for
description of the 2DEG in a perpendicular magnetic field.\cite{dz83,di05,dizh05} \vspace{-3mm}
When acting on the vacuum $|{\rm 0}\rangle$ (in our case $|{\rm
0}\rangle\!=\!|\overbrace{\uparrow,\uparrow,...\uparrow}^{\displaystyle{\vspace{-5mm}
\mbox{\tiny{$\;N_\phi$}}^{\vphantom{\int^\infty}}}}\,\rangle$), the exciton operators produce a
set of basis states which diagonalize the single-particle term of the Hamiltonian and some part
${\hat H}_{\rm ED}$ of the interaction Hamiltonian.\cite{dizh05,di02} Exciton states are
classified by ${\bf q}$, and it is essential that in this basis {\it the LL degeneracy is
lifted}.

So, the generic Hamiltonian is ${\hat H}\!=\!{\hat H}_1\!+\!{\hat
H}_{\rm int}$ where
$$\begin{array}{l}
 {\hat H}_1=\sum\limits_\sigma\int\! d{\bf r}\,{\hat \Psi}_\sigma^\dag({\bf r})
 \left[\frac{1}{2m^*}\left(
 i{\vec \nabla}-{e\vec A/c}\right)^2\!+\!g\mu_BB{\hat S}_z\right]\!{\hat \Psi}_\sigma({\bf r})\quad
 \mbox{and}\\
{\hat H}_{\rm int}=\frac{1}{2}\sum\limits_{\sigma_1,\sigma_2}\int\!
d{\bf r}_1d{\bf r}_2\,{\hat \Psi}_{\sigma_2}^\dag({\bf r}_2){\hat
\Psi}_{\sigma_1}^\dag({\bf r}_1)U({\bf r}_1\!-\!{\bf r}_2){\hat
\Psi}_{\sigma_1}({\bf r}_1){\hat \Psi}_{\sigma_2}({\bf r}_2).
\end{array}   \eqno(1)
$$
Choosing, e.g, the Landau gauge and substituting for the
Schr\"odinger operator ${\hat
\Psi}^\dag_\sigma\!\!=\!\!\sum_{np}a^\dag_{np\sigma}\psi_{np\sigma}^*$
(indexes $n,p,\sigma$ label the LL number, intra-LL state, and
spin sublevel), one can express the Hamiltonian (1) in terms of
combinations of various components of the density-matrix
operators.\cite{di05,dizh05,di02} These are exciton operators
defined as$\,$\cite{dz83,di05,dizh05,di02}
$$
  {{\cal Q}}_{ab{\bf q}}^{\dag}={N_{\phi}}^{-1/2}\sum_{p}\,
  e^{-iq_x p}
  b_{p+\frac{q_y}{2}}^{\dag}\,a_{p-\frac{q_y}{2}}\quad\mbox{and}\quad
  { {\cal Q}}_{ab{\bf q}}={ {\cal Q}}_{ba\,-{\bf q}}^{\dag}  \eqno(2)
$$
 and obeying the commutation algebra$\,$\cite{dizh05}
$$
\left[{\cal Q}_{cd\,{\bf q}_1}^{\dag},{\cal Q}_{ab\,{\bf
   q}_2}^{\dag}\right]\!\equiv\! N_{\phi}^{-1/2}\left( e^{-i({\bf
   q}_1\!\times{\bf q}_2)_z/2}\delta_{b,c}{\cal Q}_{ad\,{\bf q}_1\!+\!{\bf
   q}_2}^{\dag}-e^{i({\bf
   q}_1\!\times{\bf q}_2)_z/2}\delta_{a,d}
   {\cal Q}_{cb\,{\bf q}_1\!+\!{\bf
   q}_2}^{\dag}\right)                    \eqno (3)
$$
(in our units $l_B\!=\!\sqrt{c\hbar/eB}\!=\!1$). Here $a,b,c,...$ are binary indexes (see
above), which means that $a_p^\dag\!=\!a_{n_ap\,\sigma_a}^\dag$,
$b_p^\dag\!=\!{a}^\dag_{n_bp\,\sigma_b}$... We will also employ for binary indexes the notations
$n\!=\!(n,\uparrow)$ and ${\overline n}\!=\!(n,\downarrow)$, so that the single-mode component
of the CSFE is defined as ${\cal Q}^\dag_{0\overline{1}{\bf q}}|0\rangle$.  The interaction
Hamiltonian can be presented as ${\hat H}_{\rm int}\!=\!{\hat H}_{\rm ED}\!+{\hat H}'$ where
${\hat H}_{\rm ED}$, if applied to the state ${ {\cal Q}}_{ab{\bf q}}^{\dag}|0\rangle$, yields a
combination of single-exciton states with the same numbers $\delta n$, $\delta S_z$, and ${\bf
q}$ (see Refs. \onlinecite{dizh05,di02} and therein ${\hat H}_{\rm ED}$ expressed in terms of
exciton operators). In the framework of the above HF approximation, the CSFE correlation
energy$\,$\cite{pi92,lo93} is obtained from the equation ${\cal E}_{0\overline{1}}({
q})\!=\!\langle{\mbox{\rule{0pt}{4mm}}}
   0|{\cal Q}_{0\overline{1}{\bf q}}
  [{\hat H}_{\rm int},
  {\cal Q}_{0\overline{1}{\bf q}}^{\dag}]|0{\mbox{\rule{0pt}{4mm}}}
  \rangle$ where only the ${\hat H}_{\rm ED}$ part of the
interaction Hamiltonian contributes to the expectation. In the following, we need this so-called
HF value at $q=0$, namely ${\cal E}_{0\overline{1}}(0)\!\equiv\!{\cal E}_{\rm HF}
\!=\!\frac{1}{2}\int_0^{\infty}{p^3dp}V(p)e^{-p^2/2}$ where $2\pi V(q)$ is the Fourier component
of the effective Coulomb vertex in the layer.\cite{pi92} (In the strictly 2D limit $\alpha\!\to\!1$, and
$V(q)\!\to e^2/\kappa l_Bq$.)

The problem arises due to the `troublesome' part ${\hat H}'$ of
the interaction Hamiltonian which can not be diagonalized in terms
of single-exciton states. For our task we keep in ${\hat H}'$ only
the terms contributing to $\left[{\hat H}',{\cal
Q}_{0\overline{1}{\bf q}}^{\dag}\right]|0\rangle$ and besides
preserving the cyclotron part of the total energy (i.e. commuting
with ${\hat H}_1$). In terms of the ER these
are~$\,$\cite{di05,dizh05}
$$
  {\hat H}_{0{\overline 1}}'=\sum_{\bf q}\frac{q^2}{2}V(q)e^{-q^2/2}
  { {\cal Q}}_{01{\bf q}}^{\dag}
  { {\cal Q}}_{\overline{0}\,\overline{1}{\bf q}}
  \;+\;  \mbox{H.c.} \eqno (4)
$$
Using Eqs. (3) and identities  ${\cal Q}^\dag_{aa{\bf
q}}|0\rangle\!\equiv\!N_{\phi}^{-1/2}\delta_{{\bf q},{\bf
0}}|0\rangle$ if $a\!=\!(0,\uparrow)$ and ${\cal Q}^\dag_{aa{\bf
q}}|0\rangle\!\equiv\!0$ if $a\!\not=\!(0,\uparrow)$, one can find
that the operation of $ {\hat H}_{0{\overline 1}}'$ on vector
${\cal Q}_{0\overline{1}{\bf q}}^{\dag}|0\rangle$ results in a
combination of states of the type of ${N_{\phi}}^{-1/2}
  \sum_{\bf s}f({\bf s})
  {\cal Q}_{0\overline{0}\,{\bf q}/2\!-\!{\bf s}}^\dag
  {\cal Q}_{01\,{\bf q}/2\!+\!{\bf s}}^\dag|0\rangle$ with a certain regular and
  square integrable envelope function,
  $\int\!|f({\bf s})|^2d{\bf s}\!\sim \!1$. The norm of this combination
  is not small as compared to $\langle 0|{\cal Q}_{0\overline{1}{\bf
q}}{\cal Q}^\dag_{0\overline{1}{\bf q}}|0\rangle\!\equiv\! 1$, and the terms (4) must be taken
into account when calculating the CSFE energy.

On the other hand, if the set of double-exciton states $|{\bf
s},{\bf q}\rangle\!=\!{\cal Q}_{0\overline{0}\,{\bf q}/2\!-\!{\bf
s}}^{\dag}
  {\cal Q}_{01\,{\bf q}/2\!+\!{\bf s}}^{\dag}|0\rangle$ is considered, then one finds that they, first,
  are not exactly but `almost' orthogonal: $\langle {\bf q}_1,{\bf s}_1|{\bf s}_2,{\bf q}_2\rangle\!=\!
  \delta_{{\bf q}_1,{\bf q}_2}
  \left\{\delta_{{\bf s}_1,{\bf s}_2}\right\}$, where $
\left\{\delta_{{\bf s}_1,{\bf s}_2}\right\}\!\equiv\!\delta_{{\bf
s}_1,{\bf s}_2}\!-e^{i({\bf s}_1\!\times {\bf s}_2)_z}\!/N_\phi$;
and, second, $|{\bf s},{\bf q}\rangle$ satisfies the equation
$$
   \left[{\hat H}_{\rm int},{\cal Q}_{0\overline{0}\,{\bf q}/2\!-\!{\bf s}}^\dag
  {\cal Q}_{01\,{\bf q}/2\!+\!{\bf s}}^{\dag}\right]|0\rangle=\left[{\cal E}_{\rm sw}(|{\bf q}/2\!+\!{\bf s}|)+{\cal
  E}_{\rm mp}(|{\bf q}/2\!-\!{\bf s}|)\right]|{\bf s},{\bf q}\rangle+|{\tilde
  \varepsilon}\rangle,      \eqno (5)
$$
where the state $|{\tilde \varepsilon}\rangle$ has a negligibly small norm: $\langle {\tilde \varepsilon}|{\tilde
  \varepsilon}\rangle\sim E_{\rm C}^2/N_{\phi}$. Therefore
 the double-exciton state $|{\bf s},{\bf q}\rangle$
in the thermodynamic limit actually corresponds to free
noninteracting excitons: one of them is a spin exciton (spin wave)
with energy $|g\mu_BB|\!+\!{\cal E}_{\rm sw}$ where
$$
{\cal E}_{\rm sw}(q)\!=\!\int_0^{\infty}{pdp}V(p)
  e^{-p^2/2}\left[1\!-\!J_0(pq)\right],   \eqno (6)
$$
while the other is a magnetoplasmon with energy
$\hbar\omega_c\!+\!{\cal E}_{\rm mp}$ where
$$
{\cal E}_{\rm mp}(q)=\frac{q^2}{2}V(q)
  e^{-q^2/2}\!+\!
 \int_0^{\infty}\!{pdp}e^{-p^2/2}
  V(p)\left(1-\frac{p^2}{2}\right)
  \left[1\!-\!J_0(pq)\right]     \eqno(7)
$$
[$J_0$ is the Bessel function (cf. Refs. \onlinecite{by81,ka84})].

Thus we try for the CSFE state the vector $|X_{\bf q}\rangle\!=\!{\hat X}_{\bf q}|0\rangle$
where ${\hat X}_{\bf q}$ is a combined operator
$$
  {\hat X}_{\bf q}={\cal Q}^\dag_{0\overline{1}\,{\bf q}}+
   \frac{1}{\sqrt{2N_{\phi}}}
  \sum_{\bf s}\varphi_q({\bf s}){\cal Q}_{0\overline{0}\,{\bf q}/2\!-\!{\bf s}}^\dag
  {\cal Q}_{01\,{\bf q}/2\!+\!{\bf s}}^{\dag}\,. \eqno (8)
$$
Actually only a certain `antisymmetrized' part $\{\varphi_q\}$ of the envelope functions
contributes to the double-exciton combination in $|X_{\bf q}\rangle$.\cite{by83,di05,dizh05} In
our case the antisymmetry transform is $
  \{\varphi_q\}=
  \left[\varphi_q({\bf s})-\frac{1}{N_{\phi}}
  \sum_{{\bf s}'}e^{i({\bf s}\times {\bf s}')_z}
  \varphi_q({\bf s}')\right]
$.
 Such a specific
feature originates from the generic permutation antisymmetry of
the Fermi wave function of our many-electron system. We may
therefore consider only `antisymmetric' functions for which
$$
\varphi_q\!=\!\{\varphi_q\}/2\,. \eqno (9)
$$
 Our task is to find the energy of the eigenvector $|X_{\bf q}\rangle$
 and the `wave function'
$\varphi_q({\bf s})$, assuming that the latter is regular and square integrable. If $E_q$ is the
correlation part of the total CSFE energy (namely, $E_{\rm CSFE}\!=\!E_{\rm
vac}\!+\!|g\mu_BB|\!+\!\hbar\omega_c\!+\!E_q$), then $E_q$ is found from$\,$\cite{foot2}
$$
  \left[{\hat H}_{\rm ED}\!+\!{\hat
  H}_{0\overline{1}}'\,,{\hat X}_{\bf q}\right]|0\rangle=E_q|X_{\bf q}\rangle. \eqno (10)
$$
Now we project this equation onto two basis states $|{\bf p},{\bf q}\rangle$ and ${\cal
Q}^\dag_{0\overline{1}{\bf q}}|0\rangle$, and obtain two closed coupled equations
$$
\begin{array}{l}
  \left(2N_{\phi}\right)^{1/2}\left\langle{\bf q},{\bf p}|\left[{\hat H}_{01}',
  {\cal Q}^\dag_{0\overline{1}{\bf 0q}}\right]
  |0\right\rangle\qquad{}\qquad{}\\
  {}\qquad{}\qquad+\sum\limits_{\bf s}\varphi_q({\bf s})\left\langle{\bf q},{\bf p}| \left[{\hat H}_{ED},{\cal
Q}_{0\overline{0}\,\!{\bf q}/2\!-\!{\bf s}}^\dag
  {\cal Q}_{01\,\!{\bf q}/2\!+\!{\bf s}}^\dag\right]|0\right\rangle=E_q\varphi_q({\bf p})
\end{array} \eqno (11)
$$
and \vspace{-2mm}
$$
  {\cal E}_{0\overline{1}}(q)+(2N_{\phi})^{-1/2}\sum_{\bf s}\varphi_q
  ({\bf s})\left\langle 0|{\cal Q}_{0\overline{1}{\bf q}}\left
  [{\hat H}_{01}',{\cal Q}_{0\overline{0}\,\!{\bf q}/2\!-\!{\bf s}}^\dag
{\cal Q}_{01\,{\bf q}/2\!+\!{\bf s}}^\dag\right]|0\right\rangle=E_q
\eqno (12)
$$
for $E_q$ and $\varphi_q({\bf p})$.

Next step is a routine treatment of Eqs. (11) and (12) in terms of calculation of commutators
guided by commutation rules (3). In the ${\bf q}\!=\!0$ case, which we immediately consider, the
function $\varphi_0({\bf p})$ depends only on the modulus of ${\bf p}$. As a result we obtain$\,$\cite{foot3}
$$
  \begin{array}{r}
  \left[E-{\cal E}_{\rm sw}(q)-{\cal E}_{\rm mp}(q)\right]\varphi_0(q)+
  \displaystyle{\int_0^{\infty}\!\! s ds}
  \left[K_1(s,q)\varphi_0(s)\vphantom{\displaystyle{\frac{K_2(s)}{\pi}\!\int_0^{\pi}}}\right.
  \quad{}\qquad{}\qquad{}\\
  {}\qquad{}\qquad{}\qquad{}+\left.\displaystyle{\frac{K_2(s)}{\pi}\!\int_0^{\pi}\!
  d\phi}\left(1\!-\!\cos[{\bf s}\times{\bf q}]\right)\varphi_0(|{\bf q}\!+\!{\bf s}|)\right]=g(q)
  \end{array}                \eqno (13)
$$
\vspace{-3mm} and
$$
  E-{\cal E}_{\rm HF}=\frac{1}{\sqrt{2}}\int_0^{\infty}\!\!dpp^3V(p)
  e^{-p^2/2}\varphi_0(p)  \eqno (14)
$$
(we omit subscript 0 in $E_0$), where
$$
  g(q)=\frac{q^2}{2\sqrt{2}}V(q)e^{-q^2/2}-
  \frac{1}{2\sqrt{2}}\int_0^{\infty}p^3V(p)e^{-p^2/2}J_0(pq)dp, \eqno(15)
$$
\vspace{-3mm}
$$
  K_1(q,s)=\frac{s^2}{2}e^{-s^2/2}V(s)J_0(qs),\quad\mbox{and}\quad{}
   K_2(s)=\left(2\!-\!\frac{s^2}{2}\right)V(s)e^{-s^2/2}
   \eqno(16) \vspace{-2mm}
$$
($\phi$ in Eq. (13) is the angle between ${\bf s}$ and ${\bf q}$).

The problem has thus been integrable to yield in the thermodynamic limit a pair of coupled
integral equations for one-dimensional function $\varphi_0(q)$ and the eigenvalue $E$. In order
to solve this system we employ the method of expansion in orthogonal functions
$$
  \varphi_0(q)=\sum_{n=1,3,5,...}^{2N-1}A_n\psi_n(q)\,.
  \eqno (17)
$$
These $\psi_n\!=\!\sqrt{2}L_n(q^2)e^{-q^2/2}$ with odd indexes of the Laguerre polynomials
($\int_0^\infty qdq\psi_m\psi_n\!=\!\delta_{m,n}$) are chosen as a natural basis satisfying: (i)
the property of integrability and expected analytic and asymptotic features of $\varphi_0(q)$;
(ii) the antisymmetry condition (9). In other words, we change from the basis formed by the set
of nonorthogonal double-exciton states $|{\bf s},{0}\rangle\!\equiv\!{\cal
Q}_{0\overline{0}-\!{\bf s}}^\dag {\cal Q}_{01{\bf s}}^\dag|0\rangle$ to a new set of basis
states $|{\rm DX},n\rangle\!=\!(2N_\phi)^{-1/2}\sum_{\bf s}\psi_n(s)|{\bf s},{ 0}\rangle$ which
are strictly orthogonal. Indeed, one can check by employing Eq. (3) and identity
$\frac{1}{N_\phi}\sum_{\bf s}e^{i({\bf q}\times{\bf s})_z}\psi_n(q)\!\equiv\! \int_0^\infty
sdsJ_0(qs)\psi_n(s)\!\equiv\!-\psi_n(q)$ that $\langle m,{\rm DX}|{\rm
DX},n\rangle\!\equiv\!\delta_{m,n}$. The integer number $N$ is dimensionality of this new
double-exciton basis.

After substitution of Eq. (17) into Eq. (14) the latter  takes the
form: $E=F$, where
$$
F={\cal E}_{\rm HF}+\frac{1}{\sqrt{2}}\sum_{n=1,3,5,...}^{2N-1} A_n\int_0^{\infty}dpp^3V(p)e^{-p^2/2}\psi_n(p).   \eqno (18)
$$

Let us consider the ideal 2D case where $V(q)=1/q$. (Here and below energy is measured in units
of $e^2/\kappa l_B$.) After substitution of the expansion (18) into Eq. (13), further
multiplication by basis functions $\psi_m(q)$ and integration ($\int...qdq$) lead to the set of
$N$ linear algebraic equations with respect to $A_n$. Finding $A_n$ for a given $E$ and
substituting them into Eq. (18), we obtain $F(E)$. The condition $F(E)\!=\!E$ yields the desired
result $E=E_{\rm SF}$.

\begin{figure}[h]
\begin{center} \vspace{-8.mm}
\includegraphics*[width=0.6\textwidth]{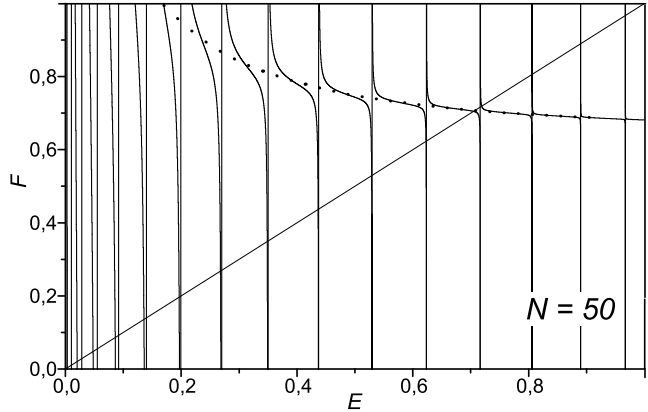}
\end{center}
\vspace{-15.mm}
 \caption{Graphical solution of Eqs. (13) and (14). Intersection of the $F\!=\!E$ straight
 line with the dotted line corresponds to the CSFE energy, $E_{\rm SF}\!\approx 0.71\!$. See
 text for details. }\vspace{-1.5mm}
\end{figure}

Fig. 1 shows the result of calculations for $N=50$. The lines which are restricted by vertical
asymptotes reflect the result of calculation of $F(E)$. Points of singularity $E^{(i)}$, at
which $F$ goes to infinity, are roots of the equation $D_N(E)=0$ where $D_N$ is the determinant
corresponding to the ``left-side'' of the set of equations for $A_n$. By increasing $N$ we
increase the order of equation $D_N(E)=0$, so that this has up to $N$ real roots. Indeed, when
observing the evolution of $F(E)$ with increasing $N$, one finds that the number of singular
points grows, and they become more densely placed. For $N\to\infty$ one could expect that a
singular point appears within an arbitrarily small vicinity of every value $E$. Since all the
vertical asymptotes $E=E^{(i)}$ are crossed by the straight line $F=E$ (see Fig. 1), we come to
the conclusion that for {\it any} $E$ there is a singular solution of Eqs. (13) and (14). Such
solutions with singular functions $\varphi(q)$ form a band. The physical meaning of this result
is quite transparent. Namely, the band corresponds to energy ${\cal E}_{\rm sw}(q)\!+\!{\cal
E}_{\rm mp}(q)$ of unbound exciton pairs. Now we only consider the solution $E=F(E)$, where the
$F\!=\!E$ line crosses a conventional envelope curve tracing {\it the regions of regularity} of
$\varphi_0$ determined by Eq. (17). Such regions at a finite $N$ should be as distant as
possible from the points of singularity, and we simply define them as the vicinities of
``middle'' points $\overline{E}^{(i)}=\frac{1}{2}(E^{(i)}+E^{(i+1)})$. The envelope curve may
obviously be defined as the line passing through the points $[\overline{E}^{(i)},\,
F(\overline{E}^{(i)})]$. The intersection with the straight line $F=E$ occurs at the only point
stable with respect to evolution of this picture at $N\to\infty$. This intersection point is
readily seen in Fig. 1.

Fig. 1 shows the build-up of singular points (vertical lines) with vanishing $E$ and vice versa
a certain rarefication of singularities in the vicinity of $E_{\rm SF}$. The former reflects
growth of the density of states at the bottom of the exciton band whereas the latter is a usual
effect of the ``levels' repulsion''. Note that  the non-Hartree-Fock shift for the CSFE level is
positive as compared to the value $E_{\rm HF}\!=\!0.627$. This is expected because the repulsion
of the CSFE from the lower-lying crowded states of unbound excitons should be stronger than from
the upper states having comparatively low density. At the same time, one can also see in Fig. 1
some trend towards the concentration of singularity points $E^{(i)}$ at higher energies $E$.
This is evidently a consequence of the density of states growth at the top of the exciton band.

In general, the larger is $N$ the more accurate is the calculation of $\varphi_0(q)$ and $E$,
i.e. the envelope curve in Fig. 1 becomes discernible and may be drawn only at considerable $N$.
At the same time the analysis reveals that the intersection point with the $F\!=\!E$ line is
rather stable and only weakly depends on $N$. This feature prompts us to consider the case
$N\!=\!1$ where double-exciton states mixed with ${\cal Q}^\dag_{0\overline{1}{\bf q}}|0\rangle$
are modelled by a single vector $|{\rm DX},1\rangle$. Actually the $N\!=\!1$ approximation for
the problem determined by Eqs. (13), (14) and (17) is equivalent to a variational procedure for
the trial double-mode state $|{ X}_0^{\rm DM}\rangle\!=\!{\cal Q}^\dag_{0\overline{1}{\bf
0}}|0\rangle\!+\!A_1|{\rm DX},1\rangle$, where the correlation part of the excitation energy is
found from equation
$$
  E=\min\limits_{A_1}\!\left(\frac{\langle{ X}_0^{\rm DM}|{\hat H}_{\rm int}|{ X}_0^{\rm DM}\rangle}{\langle{ X}_0^{\rm DM}|
  { X}_0^{\rm DM}\rangle}\right)\!-\!{E}_{\rm vac}^{\rm int}\,  \eqno (19)
$$
(${E}_{\rm vac}^{\rm int}$ denotes the correlation part of the ground-state energy). After minor
manipulations we find that this simple double-mode approximation (DMA) reduces our problem to
the secular equation $\mbox{det}\!\left|(E-{\EuScript
E}_i)\delta_{ik}+(1\!-\!\delta_{ik}){\EuScript D}_{ik}\right|\!=\!0$ (indexes $i$ and $k$ are 1
or 2), where ${\EuScript E}_1\!=\!\int_0^\infty\!qdqV(q)\epsilon(q),\;$ $ {\EuScript
E}_2\!=\!{\cal E}_{\rm HF},\;$ and ${}\;{\EuScript D}_{12}\!\equiv\!{\EuScript
D}_{21}\!=\!\int_0^\infty\!qdqV(q)d(q)$  with $ \epsilon\!=\!2q^2(1\!-\!q^2)^2e^{-3q^2/2}\!+\!
\frac{1}{2}(q^2\!-\!5q^2\!+\!q^4)e^{-q^2}\!
-\!\frac{1}{16}(q^2\!-\!4)^3e^{-3q^2/4}\!+\!({q^2}/{2}\!-\!2)e^{-q^2/2}$
{and} $d\!=\!q^2(q^2\!-\!1)e^{-q^2}.$ Only the largest root of this secular equation has
physical meaning. In the ideal 2D case we easily obtain the DMA correlation energy of the CSFE:
$E_{\rm SF}= 0.766$. Comparing this result with Fig. 1
we conclude that even the DMA works rather well.

\begin{figure}[h]
\begin{center} \vspace{-8.mm}
\includegraphics*[width=0.6\textwidth]{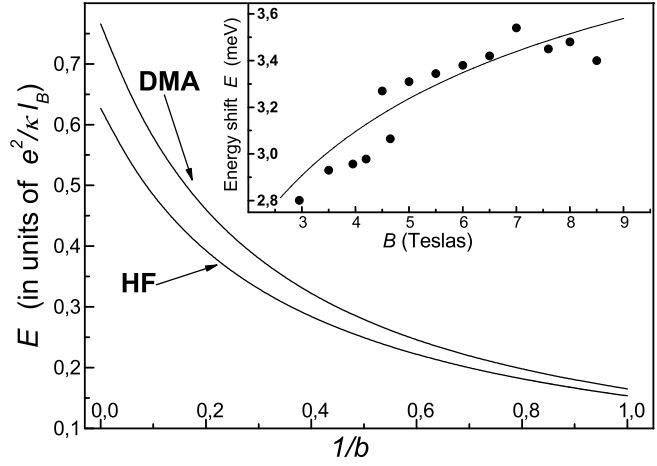}
\end{center}
\vspace{-16.mm}
 \caption{Main picture: DMA and HF shifts in dimensionless units against the form-factor
 parameter $b$. Inset: DMA shift against the magnetic field when $b\!=\!5.45\,B^{-1/2}$
 ($b\!=\!0.213\,l_B\!/\mbox{nm},\;$ $l_B$ in nm's, $B$ in Teslas); symbols are experimental data
  for the $25\,$nm quantum wells.\cite{va06} }\vspace{-1.5mm}
\end{figure}

Fig. 2 shows the CSFE correlation energy calculated within the DMA and employing the HF approximation, if
the vertex $V$ for a real 2DEG is defined as $V\!=\!F_b(q)/q$ with
the formfactor $
  F_b(q)\!=\!\frac{1}{8}\left(1\!+\!\frac{q}{b}\right)^{-3}\left[8\!+\!9
  \frac{q}{b}\!+\!3\left(\frac{q}{b}\right)^2\right].$\cite{an82,lo93}
Here $b=b_0l_B$ is a dimensionless parameter corresponding to dimensionless $q$. ($b_0$ is
considered to be independent of the magnetic field.) It is seen that the non-HF shift of the
CSFE energy, being about $15\%$ in the strict 2D limit (i.e., in the $b\!\to\!\infty$ case),
becomes smaller ($\sim\! 5-6\%$) in real samples. This difference is not observable
experimentally.\cite{va06} Meanwhile, the DMA results are in good agreement with experimental
data where the CSFE correlation energy is measured as a function of magnetic field, see inset in
Fig. 2. The chosen value, $b_0\!=\!0.213/$nm, is quite consistent with the available wide
quantum wells.\cite{va06}

In conclusion, we note that preliminary analysis indicates that the non-HF shift should be more substantial in the case of a fractional filling, e.g. at $\nu\!=\!1/3$. Moreover, contrary to the single-mode approximation$\,$\cite{lo93}
shifting the energy to lower values as compared to the HF result, the approach taking into account the double-exciton component should lead to a considerable positive shift in the CSFE correlation energy.

The authors acknowledge support of the RFBR and hospitality of the Max Planck Institute for Physics of Complex Systems (Dresden) where partly this work was carried out. The authors also thank I.V. Kukushkin, L.V. Kulik and A.B. Van'kov for the discussion.

\end{document}